\documentclass[epj]{svjour}
\usepackage{amssymb,amsmath}
\usepackage{graphicx}

\newcommand{\dd}                { \mathrm{d}}        
\newcommand{\ic}                { \mathrm{i}}        
\newcommand{\e}                 { \mathrm{e}}        
\newcommand{\abs}[1]            { \left|#1\right|}   
\newcommand{\V}[1]              { \mathbf{#1}}       
\newcommand{\M}[1]              { \mathbf{#1}}       
\newcommand{\trans}[1]          { {#1}^\mathrm{T}}   
\newcommand{\trace}             { \mathrm{tr}}       
\newcommand{\sfrac}[2]        { {\textstyle \frac{#1}{#2}} } 

\newcommand{\w}                 { \omega}
\newcommand{\wt}                { \tilde{\omega}}
\newcommand{\wo}                { \omega_{0}}

\newlength{\cwidth}
\newlength{\twidth}

\begin{document}
\title{Dielectric resonances of ordered passive arrays}
\author{Steffen Sch\"afer, Laurent Raymond and Gilbert Albinet}
\authorrunning{Sch\"afer {\sl et al.}}
\titlerunning{Dielectric resonances of ordered passive arrays}
\institute{Laboratoire Mat\'eriaux et Micro\'electronique de
Provence (UMR CNRS 6137) and Universit\'e Aix-Marseille I \\
B\^at.~IRPHE, Technopole de Ch\^ateau-Gombert, 49 rue Joliot Curie,
B.P.~146, 13384 Marseille Cedex 13, France.}
\offprints{Steffen Sch\"afer \email{steffen.schaefer@l2mp.fr}} 
\mail{same address}
\date{September 7, 2004}

\abstract{
The electrical and optical properties of ordered passive arrays,
constituted of inductive and capacitive components, are usually
deduced from Kirchhoff's rules.
Under the assumption of periodic boundary conditions, comparable
results may be obtained via an approach employing transfer
matrices. In particular, resonances in the dielectric spectrum are
demonstrated to occur if all eigenvalues of the transfer matrix of the
entire array are unity. The latter condition, which is shown to be
equivalent to the habitual definition of a resonance in impedance for
an array between electrodes, allows for a convenient and accurate
determination of the resonance frequencies, and may thus be used as a
tool for the design of materials with a specific dielectric response.
For the opposite case of linear arrays in a large network, where
periodic boundary condition do not apply, several asymptotic
properties are derived.
Throughout the article, the derived analytic results are compared to
numerical models, based on either Exact Numerical Renormalisation or
the spectral method.
}
\maketitle
 
\section{\label{sec:intro}Introduction}

Recent experimental advances \cite{Pendry2004} have put to living the
speculation about materials with negative refraction index, initiated
in 1968 by Victor Veselago on purely theoretical grounds
\cite{Veselago1968}.  Such materials, starting to be at reach
nowadays primarily for the microwave region, manifest exciting and
unconventional phenomena ranging from the Inverse Doppler effect
\cite{Seddon2003} to the possibility of diffraction-free imaging
\cite{Grbic2004}. 

Were such ``left-handed'' materials, as Veselago called them, available
for any frequency domain, they would undoubtedly revolutionise optics.
However, a negative refraction index requires simultaneously a
negative permittivity and a negative permeability, implying that the
material has dielectric and magnetic resonances in the same frequency
domain --- a property which turns out to be extremely rare and which is
partly responsible why such materials have remained undiscovered for
almost three decades.

Unlike to the optical domain, where the electric and magnetic
resonances are generally separated in frequency by several orders of
magnitude, negatively refracting materials have recently been
engineered for narrow bands in the microwave domain. One of the main
ingredients of these ``metamaterials'' are regular arrays of so-called
split-ring resonators \cite{Pendry2004}.

In this paper, we will focus on the dielectric part of the response of
arrays of similar resonators in square lattices. One goal is to
provide tools for the location of the resonances in the spectrum, a
task which cannot simply be performed by looking at the building
blocks of such an array \cite{Laurent2003}. At the same time, the
theoretical approaches presented in this paper allow for a deeper
understanding of the resonance spectrum --- certainly an advantage in
the quest of materials with a tailored electromagnetic
response. Throughout the paper, the analytical results are compared to
numerically calculated spectra, obtained either via Exact Numerical
Renormalisation (ENR) \cite{Vinograd,Sarych,Tort} or from the spectral
method \cite{Straley,Bergman,Milton} formulated with Green's functions
\cite{Clerc1996,Luck1998,Laurent2000}.

\begin{figure}[htb]
\begin{center}
\includegraphics[width=0.5\cwidth]{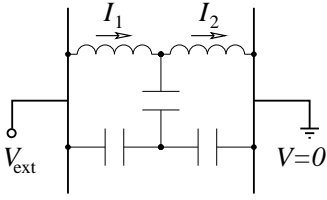}
\end{center}
\caption{\label{fig:simple} One of the simplest non-trivial clusters,
consisting of two self-induction coils embedded in a capacitor
network.} 
\end{figure}

An archetype of the circuits to be dealt with in this paper is
depicted in Fig.~\ref{fig:simple}. 
This ``cluster'' consists of two coils in series (each of inductance
$L$ and with no ohmic resistance, $R=0$) embedded in the simplest
imaginable network made of three capacitors (each of capacity $C$). As
will be seen in the following, this circuit contains --- despite its
simplicity --- many of the essential features of the more complicated
arrays to be studied in the following sections.

With no external voltage applied to the plates, the state of the
system can be obtained by straightforward application of Kirchhoff's
rules: the loop rule yields a differential equation for the current
through any of the coils $I_{k}$ in terms of the currents flowing
through the vertical and the horizontal capacitor belonging to the
same loop; the latter currents are subsequently eliminated by
Kirchhoff's junction rule, resulting in a homogeneous system of two
ordinary differential equations for the currents through the coils.
Applying a time-dependent external voltage $V_{\rm ext}(t)$ to the
plates introduces an inhomogeneity in the equations, which finally
read
\begin{equation}
\label{simple:ode}
\frac{1}{\wo^2} \frac{\dd^2}{\dd t^2} 
\begin{pmatrix} I_{1} \\ I_{2} \end{pmatrix}
\,-\,
\begin{pmatrix} -2 & \;1 \\ \;1 & -2 \end{pmatrix}
\begin{pmatrix} I_{1} \\ I_{2} \end{pmatrix}
\;=\;
I_{\rm ext}(t) \begin{pmatrix} 1 \\ 1 \end{pmatrix}
\text{,}
\end{equation} 
with $\wo=1/\sqrt{LC}$ and  $I_{\rm ext}(t)$ the total net current
flowing through the sample. 

Eq.~(\ref{simple:ode}) can be solved by standard means yielding the
homogeneous eigenmodes
\begin{equation}
\label{simple:Ih}
\V{I}_{\rm h}(t)
\;=\; 
c_{1} \begin{pmatrix} 1 \\ 1 \end{pmatrix} \e^{\ic\wo t}
\,+\,
c_{2} \begin{pmatrix} \;1 \\ -1 \end{pmatrix} \e^{\ic \sqrt{3}\wo t}
\;\mbox{,}
\end{equation}
where $c_{1}$ and $c_{2}$ are arbitrary complex amplitudes. The
application of a sinusoidal external voltage of frequency $\Omega$,
implying an external current $I_{\rm ext}(t)=\hat{I}_{\rm
ext}\exp(\ic\Omega t)$, allows for {\em one} additional solution
\begin{equation}
\label{simple:Iinh}
\V{I}_{\rm inh}(t)
\;=\;
I_{\rm ext}(t) \begin{pmatrix} 1 \\ 1 \end{pmatrix} \,
\begin{cases}
\frac{\wo^{2}}{\wo^{2}-\Omega^{2}} 
& \text{for $\Omega\neq\wo$, } \\
-\frac{\ic}{2} \wo t 
& \text{for $\Omega=\wo$.} 
\end{cases}
\end{equation}

Four observations can be made at this point: 
(i) in networks consisting of two species of components only, the
state of the system is fully described by a set of differential
equations for the currents flowing through the minority components (in
our case the coils). The contribution of the majority components
(here, the capacitors) to the Kirchhoff rules is purely algebraic, and
may be eliminated from the system of equations.
(ii) The external voltage can only excite the first resonance, of
frequency $\wo$. The second resonance, $\sqrt{3}\wo$, is antisymmetric
($I_{1}=-I_{2}$) and thus orthogonal to the currents which are always
induced symmetrically in the present configuration of the voltage
plates.
(iii) The applied voltage equals the sum of the two voltage drops in
the coils, {\sl i.e.} $V_{\rm ext}=L \frac{\dd}{\dd
t}(I_{1}+I_{2})$. Off resonance, this yields a total impedance of
$Z={V_{\rm ext}}/{I_{\rm ext}}=2\ic L\Omega
\wo^2/(\wo^{2}-\Omega^{2})$ between the plates. This result may be
generalised: as a resonance is approached, the {\em internal} currents
of the sample diverge for finite applied voltage. This may be
associated with an infinite {\em conductivity} of the sample's
(internal) components. At the same time, however, a divergent
impedance and thus zero conductivity is measured between the plates.
(iv) In reality, the coils have a small but finite ohmic resistance
$R$. The eigenfrequencies are thus shifted slightly in the complex
plane towards small positive imaginary parts which damp out the
sample's free modes, eq.~(\ref{simple:Ih}). In addition to its
imaginary part, the impedance $Z$ acquires a real part consisting of a
narrow lorentzian peak centred at the resonance frequency.

The paper is organised as follows: in Sec.~\ref{sec:kirchhoff}, the
resonance spectra of regular arrays with not too complicated unit
cells are calculated directly from the solution of Kirchhoff's
rules. In Sec.~\ref{sec:transfer}, more complicated arrays are tackled
with an approach based on transfer matrices. Regular one-dimensional
(1D) arrays in an infinite two-dimensional (2D) lattice are examined
in Sec.~\ref{sec:inflattice}, and the results compared to the formerly
discussed 2D clusters. Finally, the Appendix~\ref{app:dipole} is
devoted to a simple physical approximation, based on a dipole scenario
in a 2D environment, which turns out to be helpful for the
interpretation of the spectra of linear clusters.

\section{\label{sec:kirchhoff}Direct solution of Kirchhoff's rules}

In this section, we will tackle simple regular 1D and 2D binary arrays
by a direct solution of Kirchhoff's rules. The results are then to be
compared to those obtained by Exact Numerical Renormalisation (ENR)
\cite{Vinograd,Sarych,Tort}.  The strategy of the latter algorithm
consists of eliminating the network sites one by one, while
renormalising the impedance between all couples of former neighbours
of the eliminated site such that the global impedance remains
invariant, until the electrodes are connected by just one bond, which
then carries the whole network's impedance.

\begin{figure}[htb]
\begin{center}
\includegraphics[width=0.7\cwidth]{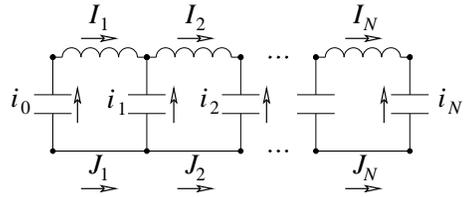}
\end{center}
\caption{\label{fig:ladderL} A ladder-shaped circuit containing $N$
coils in series on one leg, and the same number of bonds with no
impedance on the other. All $N+1$ steps of the ladder are capacitors.}
\end{figure}
To begin with, we are going to summarise the solution of Kirchhoff's
rules for the 1D array shown in Fig.~\ref{fig:ladderL}. This
ladder-shaped circuit consists of a horizontal line of $N$ coils, each
of inductance $L$, which is connected by $N+1$ capacitors, each of
capacity $C$, to another horizontal line with no resistance at
all. The loop rule states that the voltage drop around a closed loop
is zero; applying it to the $k.$th mesh and deriving with respect to
time yields
${\dd^2 I_{k}}/{\dd t^2}=\wo^{2}(i_{k}-i_{k-1})$,
with $\wo=1/\sqrt{LC}$. The currents through the capacitors, $i_{k}$,
can be eliminated using the junction rule $i_{k}+I_{k}=I_{k+1}$, where
$I_{k}$ is assumed zero if $k$ is out of range. In total, we get
\begin{equation}
\label{eqL1}
\frac{\dd^2}{\dd t^2}\V{I}\;=\;\wo^2 \M{D}\V{I}
\text{,}
\end{equation}
where $\trans{\V{I}=(I_{1},\ldots,I_{N})}$ is the coil current vector,
and $\M{D}$ the 1D lattice Laplacian, {\sl i.e.} a tridiagonal matrix
with entries $-2$ for all diagonal elements and $1$ for all non-zero
off-diagonal elements.

\begin{figure*}[htb]
\begin{center}
\includegraphics[width=0.8\twidth]{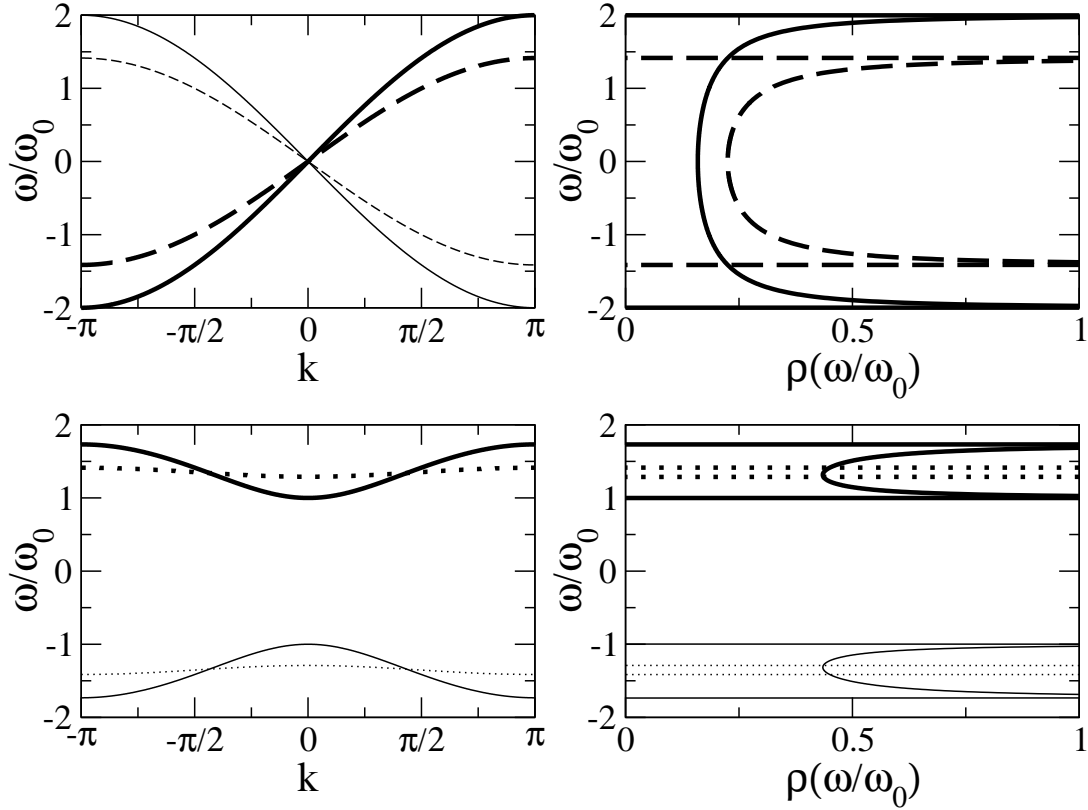}
\end{center}
\caption{\label{fig:dos} Dispersion relations $\wt(k)$ (left column)
and corresponding densities of states $\rho(\wt)$ (right column). The
two graphs in the first line correspond to the ladder of
Fig.~\ref{fig:ladderL} (continuous lines --- eqs.~(\ref{dispersL1}) and
(\ref{dosL1})) and to its 2D counterpart shown in
Fig.~\ref{fig:unitcells}(a) (dashed lines). The two graphs in the
second line where obtained for the arrays of
Fig.~\ref{fig:unitcells}(b) (continuous lines ---
eqs.~(\ref{dispersL2}) and (\ref{dosL2})), and
Fig.~\ref{fig:unitcells}(c) (dotted lines --- eq.~(\ref{dispersLC1})).}
\end{figure*}

The usual ansatz $\V{I}\sim\exp{(\ic\w t)}$ converts the differential
equation~(\ref{eqL1}) into an eigenvalue problem for the tridiagonal
matrix $\M{D}+\wt^2\M{1}$, where $\wt=\w/\wo$ stands for the frequency
in units of $\wo$. Its determinant can be calculated explicitly: for a
$N$-dimensional tridiagonal matrix $\M{M}_{N}(x)$ with diagonal
elements $2x$, and first sub- and super-diagonal elements $1$,
expansion of the determinant by minors yields
\begin{equation}
\label{chebyshev}
D_{N}(x)\;=\;2x D_{N-1}(x)\,-\,D_{N-2}(x)
\;\text{,}
\end{equation}
where $D_{N}(x)=\det{\M{M}_{N}(x)}$. We recognise in (\ref{chebyshev})
the recurrence relation generating the Chebyshev polynomials. The
required initial conditions, $D_{0}(x)=1$ and $D_{1}(x)=2x$, ties us
down to the Chebyshev polynomials of the second kind:
\begin{equation}
\label{chebyU}
U_{N}(x=\cos\theta)\;=\;\frac{\sin\left[(N+1)\theta\right]}{\sin\theta}
\end{equation}
The roots of $U_{N}(\cos\theta)$, which occur at
$\theta_{m}=\frac{m\pi}{N+1}$, determine the eigenfrequencies. In the
present example, where $2x=2\cos\theta_{m}=-2+\wt^2$, after
renumbering the eigenstates according to their frequency, we have
\begin{equation}
\label{dispersL1}
\wt_{m}=\pm 2\sin\frac{m\pi}{2(N+1)}
\quad \text{(for $m=1\ldots N$).}
\end{equation}
For infinite ladder length, the dispersion relation $\wt(k)$ {\sl
versus} wave number $k=\frac{m\pi}{N+1}$ is shown in the first graph
of Fig.~\ref{fig:dos}.
%
%
The associated density of states
\begin{equation}
\label{dosL1}
\rho(\wt)\,=\,\lim_{N\to\infty}\frac{1}{2 N}\sum\limits_{m} \delta(\wt-\wt_{m})
\,=\, \frac{1}{\pi\sqrt{4-\wt^2}}
\;\text{,}
\end{equation}
shown in the second graph, is finite for $\wt=0$, since the sample's
lowest resonance frequencies go to zero for $N\to\infty$. The fact
that $\rho(\wt)$ displays Van Hove singularities for $\wt=\pm 2$
should not be misinterpreted in the sense that the sample mainly
resonates in these frequencies when exposed to a multi-frequency
input: on the contrary, as will be pointed out in
Sec.~\ref{sec:transfer}, for reasons of symmetry there is very little
overlap between the voltage applied to the electrodes, generating an
overall current with symmetry
$\trans{\V{I}_{\rm ext}\sim (1,1,\ldots,1)}$, and the highest excited
states, with $m$ close to $N$.

\begin{figure}[htb]
\begin{center}
\includegraphics[width=0.95\cwidth]{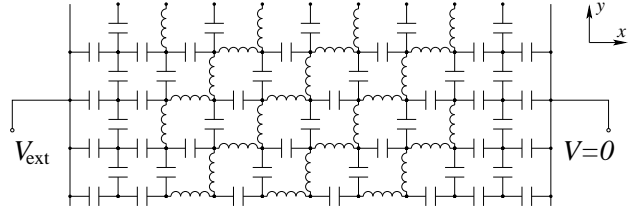}
\end{center}
\caption{\label{fig:array} Staircase-array obtained by replicating the
unit cell of Fig.~\ref{fig:unitcells}(d) $N=3$ times in
$x$-direction. The resulting structure is replicated in $y$-direction
until the electrodes are fully covered. Periodic boundary conditions
are assumed in $y$-direction.}
\end{figure}
\begin{figure}[htb]
\begin{center}
\includegraphics[width=0.9\cwidth]{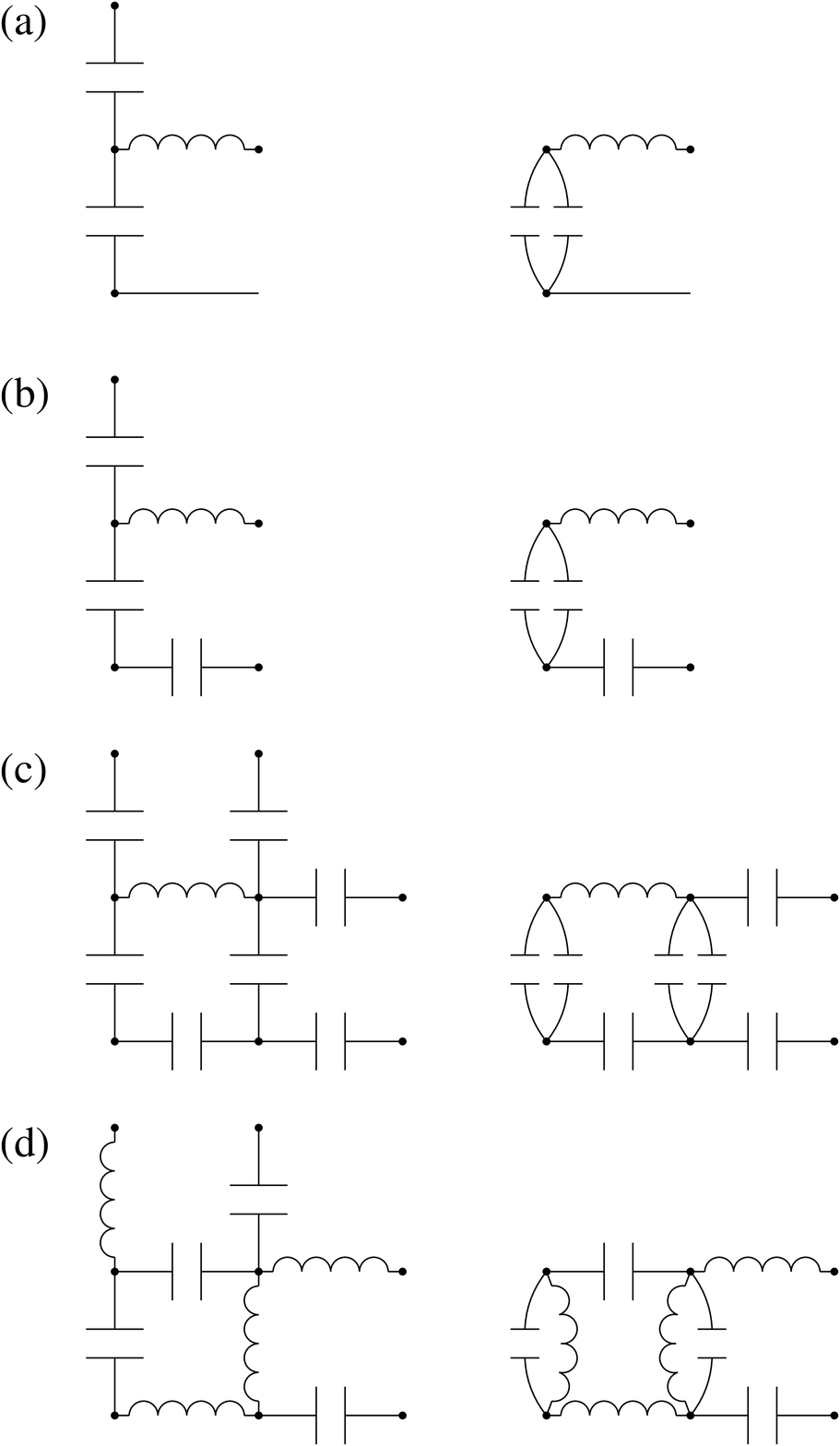}
\end{center}
\caption{\label{fig:unitcells} Unit cells for several 2D arrays (left
column) and corresponding 1D building blocks (right column).}
\end{figure}
In the following, we are going to study regular 2D arrays generated
from a unit cell which tiles the surface between the electrodes. At
first, the unit cell is replicated $N$ times in $x$-direction, {\sl
i.e.} from the left towards the right electrode. The resulting
horizontal ladder is then stacked in $y$-direction until the width of
the electrodes is fully covered, assuming periodic boundary
conditions. In other words, the 2D arrays to be studied are cylinders
of length $N$ (times the length of the unit cell). At either end, they
are connected to an electrode ring via at least one rank of pure
capacitor bonds, which avoids short circuiting and keeps the influence
of the electrodes on the dielectric spectra as low as possible.

The building block which generates an array resembling the ladder in
Fig.~\ref{fig:ladderL} is shown in Fig.~\ref{fig:unitcells}(a). 
Similar arrays can be generated from the other unit cells in
Fig.~\ref{fig:unitcells}: the array shown in Fig.~\ref{fig:array} was
obtained from the unit cell represented in Fig.~\ref{fig:unitcells}(d)
with $N=3$.

As illustrated in the right column of Fig.~\ref{fig:unitcells}, these
2D arrays can be mapped onto 1D ladders: translational invariance
along the electrodes in steps of an integer times the height of the
unit cell allows to ``bend down'' the unbound components in the top
row and to attach them to the corresponding site at the
bottom. Therefore, the only difference between the original ladder of
Fig.~\ref{fig:ladderL} and the ladder in Fig.~\ref{fig:unitcells}(a)
resumes to doubling the vertical capacitors, and thus dividing the
frequency range by a factor of $\sqrt{2}$ in eqs.~(\ref{dispersL1})
and (\ref{dosL1}). (The corresponding dispersion relation and density
of states are plotted with dashed lines in the first row of
Fig.~\ref{fig:dos}).

The array in Fig.~\ref{fig:unitcells}(b) can be tackled analogously.
The additional capacitors only contribute to the diagonal elements of
the tridiagonal matrix, and instead of eq.~(\ref{eqL1}) one obtains
$\dd^{2}\V{I}/\dd t^{2}=\frac{\wo^2}{2} [\M{D}-2\cdot\M{1}]\V{I}$. 
\begin{figure}[htb]
\begin{center}
\includegraphics[width=0.9\cwidth]{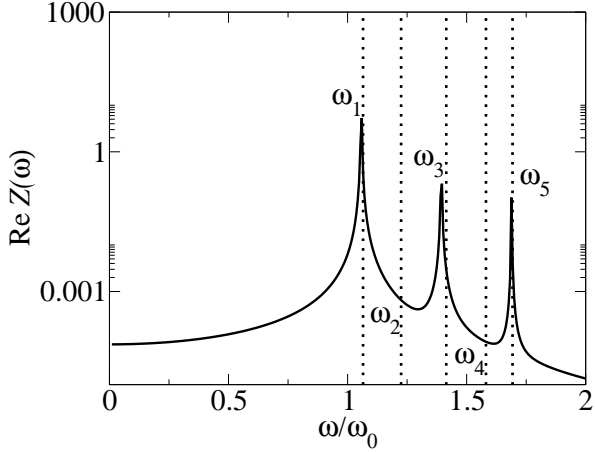}
\end{center}
\caption{\label{fig:renorm5L} 
Impedance between the electrodes for an array generated from the unit
cell in Fig.~\ref{fig:unitcells}(b). The sample consists of 6 lines,
of $N=5$ inductors in series each, substituting the central piece of
every second row in a 12 by 12 capacitor lattice. The real part of the
ENR impedance, calculated for $L=C=1$ and $R_{\rm coil}=10^{-4}$, is
plotted with continuous lines. The dotted lines indicate the resonance
frequencies obtained analytically from eq.~(\ref{dispersL2}) for $N=5$
and $m=1\ldots5$.}
\end{figure}
The resulting resonance frequencies,
\begin{equation}
\label{dispersL2}
\wt_{m}=\pm \sqrt{1+2\sin^{2}\frac{m\pi}{2(N+1)}}
\quad \text{(for $m=1\ldots N$),}
\end{equation}
are plotted, for $N=5$, with dotted lines in Fig.~\ref{fig:renorm5L}.
A corresponding ENR calculation for the total conductivity of the
array between electrodes (continuous lines in Fig.~\ref{fig:renorm5L})
detects the resonance frequencies with $m$ even, associated with
symmetric eigenmodes. The antisymmetric eigenmodes are orthogonal to
the current vector induced by the plates, and remain hence invisible
in this calculation.

The dispersion relation obtained from eq.~(\ref{dispersL2}) for
$N\to\infty$, albeit with $k=\frac{m\pi}{N+1}$ finite, along with the
corresponding density of states,
\begin{equation}
\label{dosL2}
\rho(\wt)\,=\, 
\frac{\abs{\wt}}{\pi\,\sqrt{1-\left(2-\wt^2\right)^{2}}}
\;\text{,}
\end{equation}
is illustrated by the continuous lines in the graphs in the second row
of Fig.~\ref{fig:dos}.  This time, even for infinite chain length,
there are no low-lying resonances, and a gap spreads between $-1$ and
$+1$.

The last array to be treated with this method can be obtained from a
pure capacitor lattice by replacing every second horizontal capacitor
on every second line by a coil. In this array, which corresponds to
the unit cell in Fig.~\ref{fig:unitcells}(c), meshes with one coil and
three capacitors alternate horizontally with pure capacitor
meshes. For the latter, Kirchhoff's rules reduce to purely algebraic
relations; these can be removed from the system of differential
equations, and one ends up with
\begin{equation}
\label{eqLC1}
\frac{\dd^2}{\dd t^2}\V{I}\;=\;
\frac{\wo^2}{12}
\begin{pmatrix} 
   -23 &      1 &      0 & \cdots & 0      \\
     1 &    -22 &      1 & \ddots & \vdots \\
     0 & \ddots & \ddots & \ddots & 0      \\
\vdots & \ddots &      1 &    -22 & 1      \\
     0 & \cdots &      0 &      1 & -22
\end{pmatrix}
\V{I}
\text{\;,}
\end{equation}
where $\V{I}$ is the current vector through the $N$ coils of the
cluster. The matrix in eq.~(\ref{eqLC1}) differs slightly from the
habitual form, since not all diagonal elements are the same.  The
asymmetry in the upper left corner translates the fact that the
corresponding ladder (right column of Fig.~\ref{fig:unitcells}(c))
starts with a loop of 3 capacitors and 1 coil, but ends with a
4-capacitor loop.

The solution of eq.~(\ref{eqLC1}) is analogous to the generic case,
eq.~(\ref{eqL1}), albeit leading to a slightly more complicated
recursion relation.  A cumbersome but straightforward calculation
gives the dispersion relation
\begin{equation}
\label{dispersLC1}
\wt_{m}=\pm \sqrt{\frac{5}{3}+\frac{1}{3}\sin^{2}\frac{m\pi}{2 N+1}}
\quad \text{(for $m=1\ldots N$),}
\end{equation}
with an associated density of states showing narrow bands in the range
of $\abs{\wt}=\sqrt{5/3}\simeq1.29$ to $\sqrt{2}\simeq 1.41$. As can
be seen from the dotted curves in the last two graphs of
Fig.~\ref{fig:dos}, the dispersion flattens substantially, and the
bands in the density of states are squeezed with respect to the former
case, without pure capacitor meshes (continuous lines in the same
graphs, {\sl cf.} eqs.~(\ref{dispersL2}) and (\ref{dosL2})).

This behaviour translates the tendency of the resonances to localise
on individual loops as more and more pure capacitor loops are inserted
in the ladder. In the limit of infinite distance between coils, one is
left with $N$ decoupled LC circuits of the type
\begin{equation}
\label{LCinf}
\includegraphics[height=12ex]{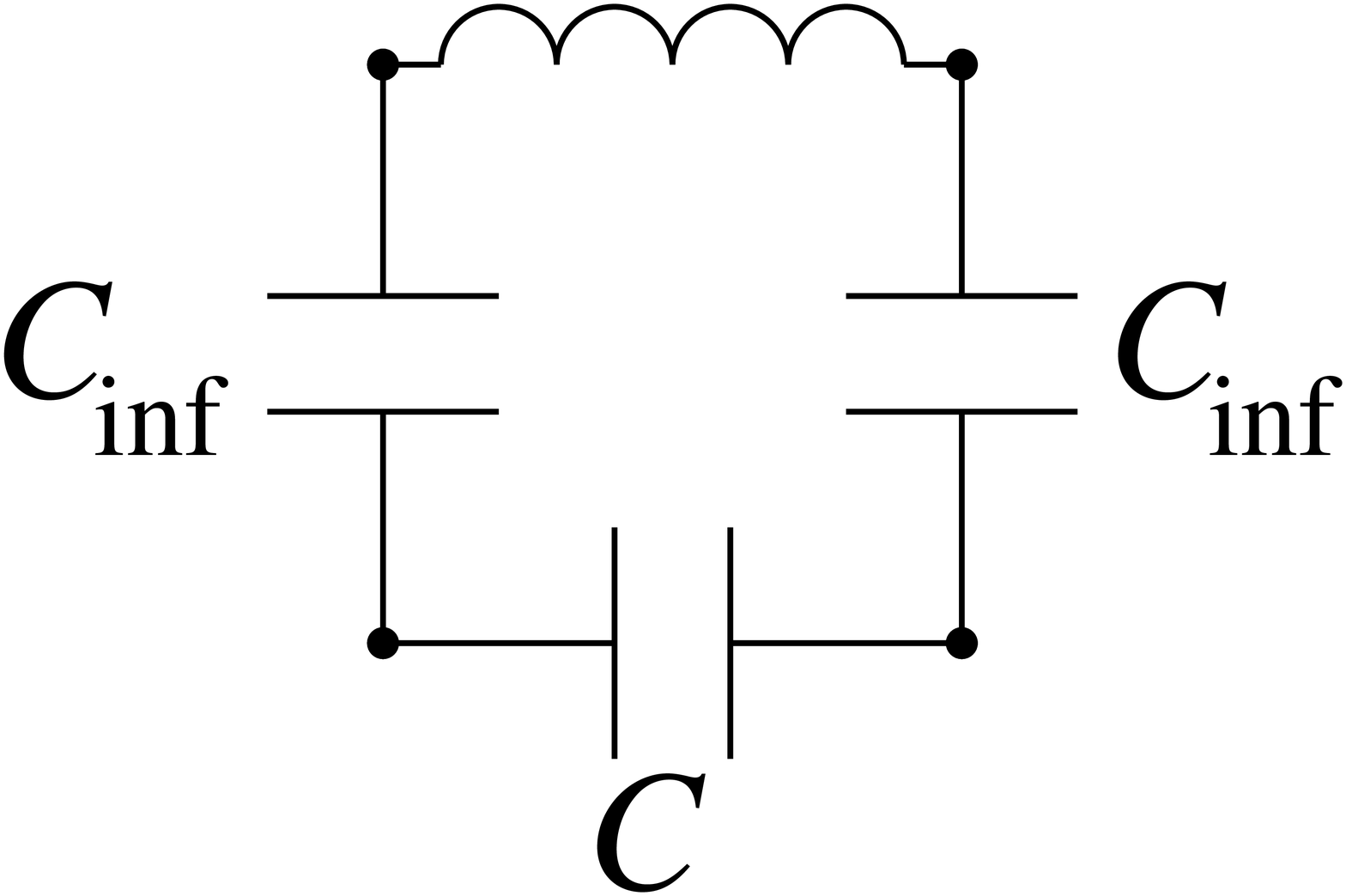}
\end{equation}
with
\begin{equation}
\label{Cinf}
\includegraphics[height=15ex]{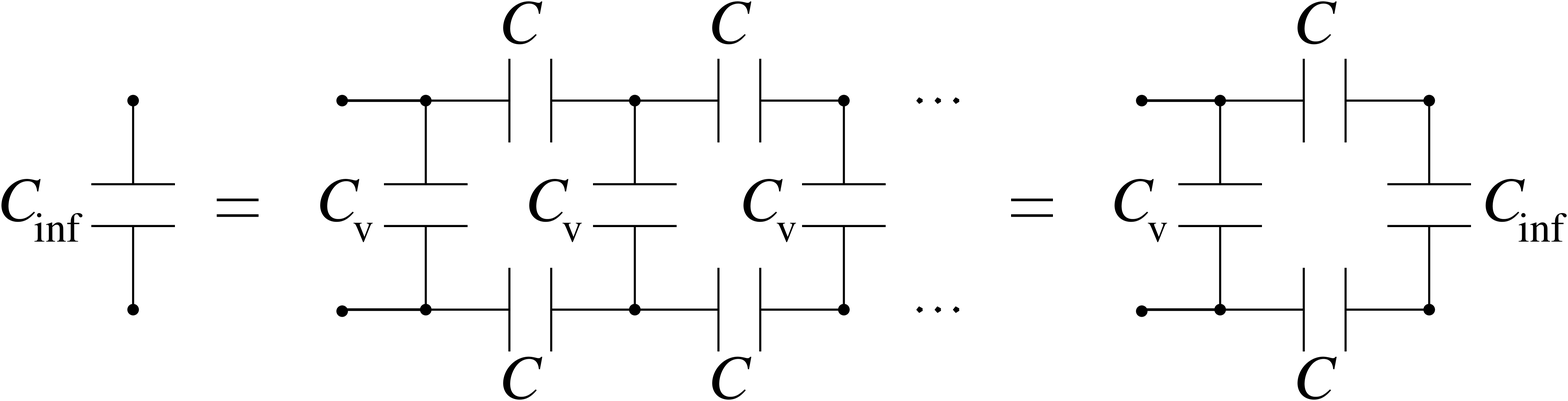}
\text{.}
\end{equation}
The solution of (\ref{Cinf}) is
$C_{\rm inf}=\frac{1}{2}C_{\rm V}(1+\sqrt{1+2C/C_{\rm V}})$
which in our case, where $C_{\rm V}=2C$, reduces to
$C_{\rm inf}=C(1+\sqrt{2})$.  
All $N$ LC circuits of type (\ref{LCinf}) resonate thus at the same
$\wt=\pm\sqrt{2\sqrt{2}-1}\simeq\pm 1.35$, and the density of
states of the system reduces to delta functions at these frequencies
which lie in the centre of the narrow dotted bands shown in the fourth
graph of Fig.~\ref{fig:dos}.

The main advantage of the direct solution of Kirchhoff's rules,
presented for several arrays in this section, are its analytical
results and the physical insight it provides into the structure of the
resonance spectra. On the other hand, for increasingly complex arrays,
the method requires --- if feasible at all --- more and more cumbersome
calculations for the solution of the recursion relations it relies
on. We will therefore present an algorithm which circumvents this
problem in the next section.

\section{\label{sec:transfer}Transfer matrix method}

The analysis of the resonance spectra of the arrays discussed so far
--- which due to their translational invariance along the electrodes
reduce to effective 1D problems --- can be rephrased very efficiently
using transfer matrices.

To illustrate the concept of a transfer matrix, consider the
quadrupole
\begin{equation}
\label{quadrupole}
\includegraphics[width=18ex]{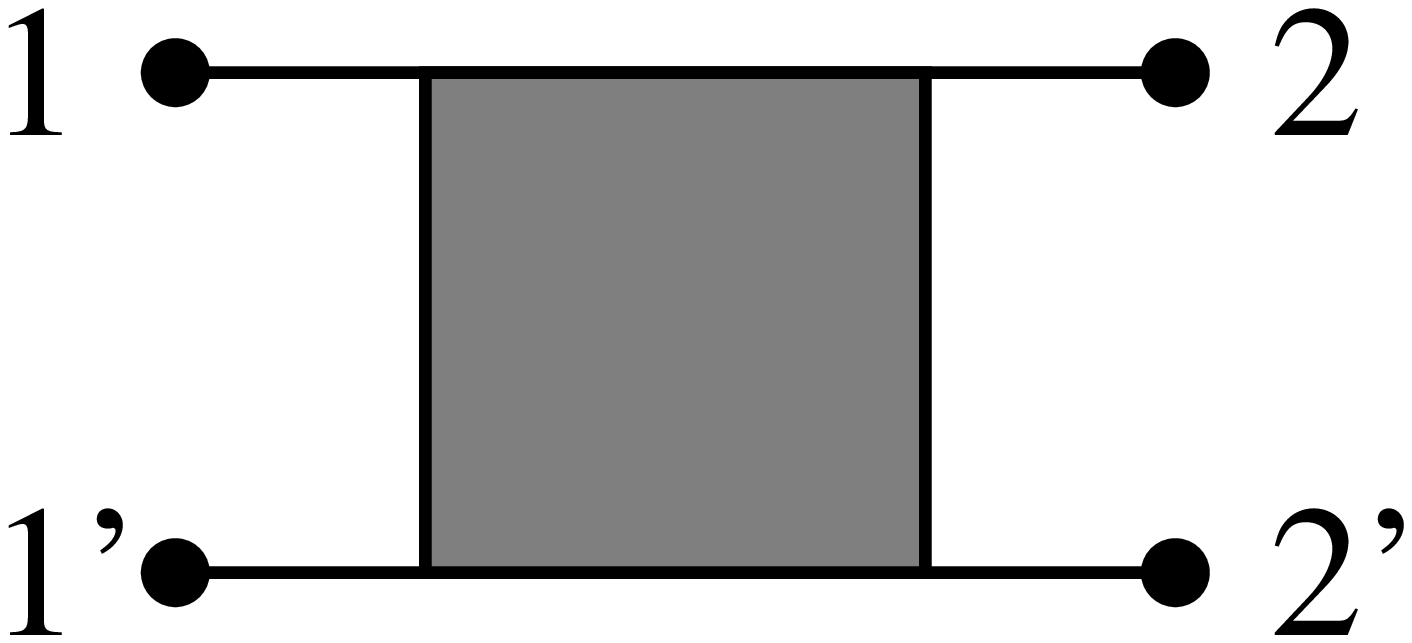}
\;\text{.}
\end{equation}
Its incoming and outgoing currents and voltages are connected to each
other via
\begin{equation}
\label{transfer}
\V{J}_{2}\;=\;\M{T}\V{J}_{1}
\;\text{,}
\end{equation}
where $\trans{\V{J}_{k}=(U_{k},I_{k},U_{k'},I_{k'})}$ is the vector of
state at each end $k=1,2$, {\sl i.e.} a combination of currents $I$
and voltages $U=V\sigma_{0}$ (the latter, for convenience, multiplied
by the conductivity $\sigma_{0}$ of the lattice's majority component);
$\M{T}$ represents the dimensionless transfer matrix.

For quadrupoles exclusively made of passive elements, there are two
conservation laws. The first states that the potentials are determined
up to an arbitrary additive constant $U_{0}$. Increasing all incoming
potentials by $U_{0}$ shifts the outgoing ones by the same amount:
hence $\trans{(1,0,1,0)}$ is a right eigenvector of the generally
non-symmetric matrix $\M{T}$, with corresponding eigenvalue
$\mu=1$. Secondly, the continuity equation states that the total
incoming and outgoing current have to be the same: hence, $(0,1,0,1)$
is a left eigenvector of $\M{T}$, again associated with $\mu=1$.

These conservation laws may be easily verified since all passive
quadrupoles can be assembled by combining two prototypes,
\begin{subequations}
\label{transferAB}
\begin{equation}
\label{transferABgraph}
\begin{tabular}{ccc}
\includegraphics[width=18ex]{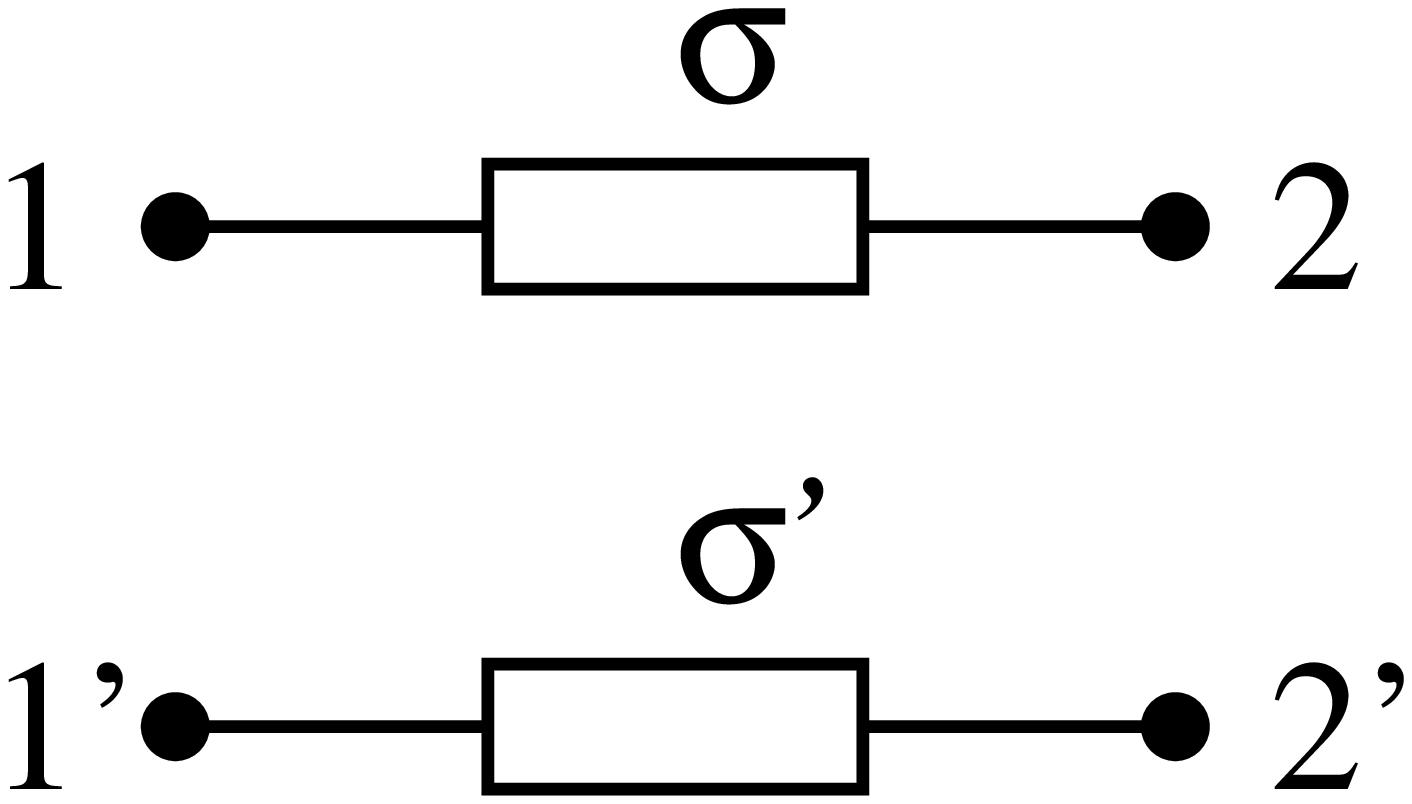} & \; & 
\includegraphics[width=18ex]{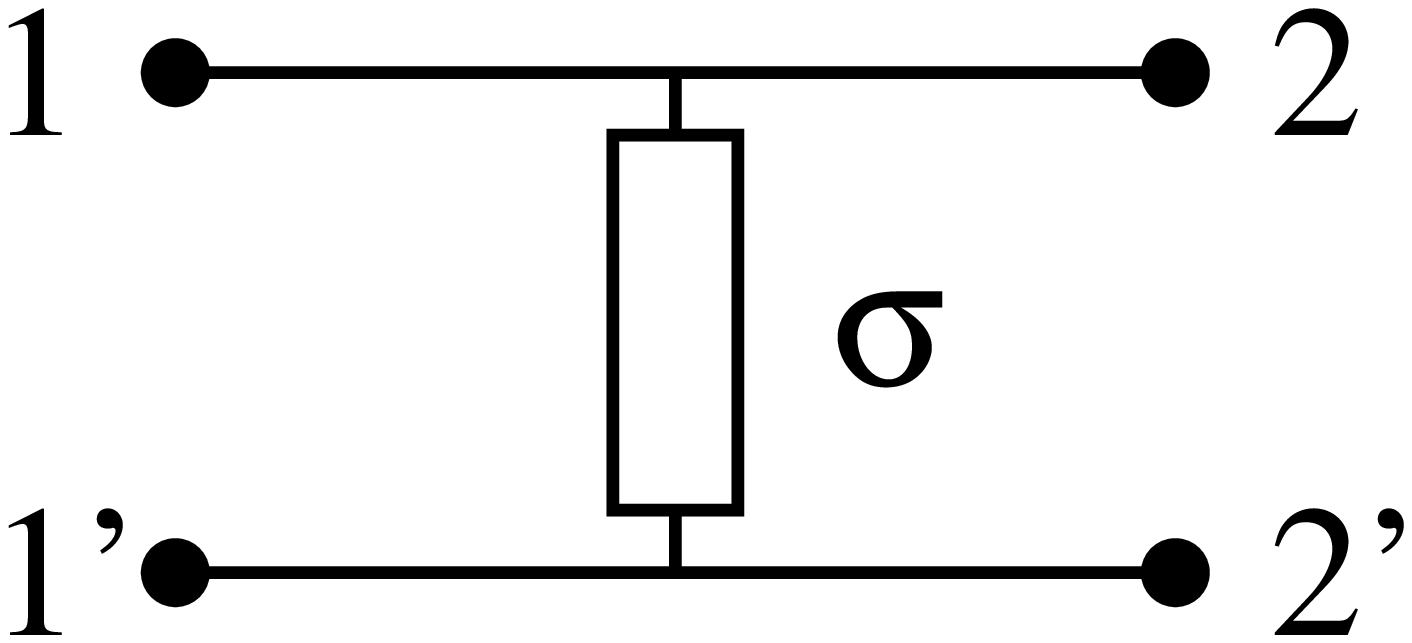} \\
\text{(A)} && \text{(B)}
\end{tabular}
\end{equation}
with corresponding transfer matrices
\begin{equation}
\label{transferABmatrix}
\M{T}_{\rm A}=
\begin{pmatrix} 
1 & -\frac{1}{\eta} & 0 & 0                \\
0 &               1 & 0 & 0                \\
0 &               0 & 1 & -\frac{1}{\eta'} \\
0 &               0 & 0 & 1                 
\end{pmatrix}
\;\text{and}\;
\M{T}_{\rm B}=
\begin{pmatrix} 
1     &  0 &     0 & 0 \\
-\eta &  1 & \eta  & 0 \\
0     &  0 &     1 & 0 \\
\eta  &  0 & -\eta & 1                 
\end{pmatrix}
\text{,}
\end{equation}
\end{subequations}
where $\eta=\sigma/\sigma_{0}$ (and $\eta'=\sigma'/\sigma_{0}$) denote
the conductivities of the components (in units of
$\sigma_{0}$). Furthermore, since the transfer matrix $\M{T}$ of any
passive quadrupole can be written as a product of matrices of type
$\M{T}_{\rm A}$ and $\M{T}_{\rm B}$, one always has $\det\M{T}=1$ ---
a relation that will be of some importance in the following.

Naively one could define a resonance as ``if what goes in, comes
out''. This amounts to resolving
\begin{equation}
\label{naivetransfer}
\begin{pmatrix}
U_{2} \\ I_{2} \\ U_{2'} \\ I_{2'} 
\end{pmatrix}
=
\begin{pmatrix}
0 \\ I_{1} \\ w \\ I_{1'} 
\end{pmatrix}
=
\M{T}
\left[
\begin{pmatrix}
0 \\ I_{1} \\ w \\ I_{1'} 
\end{pmatrix}
+\Delta U
\begin{pmatrix}
1 \\ 0 \\ 1 \\ 0
\end{pmatrix}
\right]
\;\text{,}
\end{equation}
where (arbitrarily) leg $2$ has been grounded, $U_{2}=0$; $\Delta U$
denotes the voltage drop between the two sides of the quadrupole, and
$w$ stands for the potential difference between two legs on the same
side of the quadrupole. When placed between electrodes, a resonance
would be seen as an impedance pole, implying that at finite $\Delta U$
the net current through the sample is zero,
$I_{1}+I_{1'}=I_{2}+I_{2'}=0$.  Apart from the technical difficulty
that (\ref{naivetransfer}) relies on the solution of a singular system
of equations, this method nicely yields the lowest eigenfrequency of
the quadrupole.

By contrast, the same strategy applied to a chain of $N$ identical
quadrupoles in line (with transfer matrix $\M{T}^{N}$), fails to
detect any additional eigenfrequency. To understand this failure, we
recall that eq.~(\ref{naivetransfer}) implicitly attaches each of the
two outgoing legs of the chain to the corresponding incoming leg, thus
enforcing periodic boundary conditions under which the system acquires
an additional translational symmetry along the chain. The lowest
eigenmode of the system possesses the full translational symmetry and
shows the same current distribution for each of the $N$ quadrupoles,
implying an eigenvector collinear to the current induced by the
electrodes, $\trans{\V{I}_{\rm ext}\sim (1,1,\ldots,1)}$. Since all
other eigenvectors are orthogonal to the groundstate, they cannot be
seen by simply looking ``if what goes in comes out''. (This fact is
rigorous for periodic boundary conditions; for closed boundary
conditions, as assumed in Sec.~\ref{sec:kirchhoff}, the same mechanism
gradually suppresses the response of the higher eigenmodes who have
very little overlap with the groundstate --- see
Fig.~\ref{fig:renorm5L}.)

We will now present a new resonance condition which remedies these
shortcomings; namely that, at resonance, all 4 eigenvalues $\mu$ of
the transfer matrix of the entire chain should be unity. At this
point, one may think of it as a necessary condition which prevents the
norm of the state vector $\V{J}$ from diverging or going to zero after
many transfers through the same quadrupole chain.  The proof that the
new resonance condition is equivalent to the usual definition --- {\sl
i.e.}  a divergent impedance between the electrodes (and thus zero net
current through the sample) --- is deferred to the end of this
section.

In order to simplify the following calculations, we introduce the
hermitian and unitary matrix
\begin{equation}
\label{matrixP}
\M{P}=\frac{1}{\sqrt{2}}
\begin{pmatrix}
1  &\;  0 &\;  1 &\;  0 \\
0  &\;  1 &\;  0 &\;  1 \\
1  &\;  0 &   -1 &\;  0 \\
0  &\;  1 &\;  0 & -1                 
\end{pmatrix}
\end{equation}
which allows for a transformation of the state vector to a symmetric
basis,
$\V{\tilde{J}}_{k}=\M{P}\V{J}_{k}
=\trans{(U_{k}^{+},I_{k}^{+},U_{k}^{-},I_{k}^{-})}$
with components $U_{k}^{\pm}=\frac{1}{\sqrt{2}}(U_{k}\pm U_{k'})$ and
$I_{k}^{\pm}=\frac{1}{\sqrt{2}}(I_{k}\pm I_{k'})$. The associated
transfer matrix of a single quadrupole in this basis is
$\M{S}=\M{P}\M{T}\M{P}$, and thus of the general form
\begin{math}
\M{S}=\left(
\begin{smallmatrix}
\M{A} & \M{B} \\ \M{C} & \M{D}
\end{smallmatrix}
\right)
\end{math}
with the $2\times 2$ submatrices
\begin{subequations}
\begin{align}
\label{matrixSsub}
\M{A}&=\begin{pmatrix} 1 & a \\ 0 & 1 \end{pmatrix}&
\M{B}&=\begin{pmatrix} b_1 & b_2 \\ 0 & 0 \end{pmatrix}
\\
\M{C}&=\begin{pmatrix} 0 & c_1 \\ 0 & c_2 \end{pmatrix}&
\M{D}&=\begin{pmatrix} d_1 & d_2 \\ d_3 & d_4 \end{pmatrix}
\text{.}
\end{align}
\end{subequations}
Due to the particular structure of the submatrices --- implying
$\M{A}^{n}=
\left(\begin{smallmatrix} 1 & n a \\ 0 & 1\end{smallmatrix}\right)$,
$\M{A}\M{B}=\M{B}$, $\M{C}\M{A}=\M{C}$, and $\M{C}\M{B}=\M{0}$ --- the
transfer matrix of a chain of $N$ identical quadrupoles can be
calculated explicitly; as may be proved by induction, it reads
\begin{equation}
\label{matrixSN}
\M{S}^{N}=
\begin{pmatrix}
\M{A}^{N}+\M{B}\sum\limits_{k=0}^{N-1}\M{G}_{k}\M{C} &\;& \M{B}\M{G}_{N} \\
\M{G}_{N}\M{C} && \M{D}^{N}
\end{pmatrix}
\text{,}
\end{equation}
with the geometric series 
\begin{equation}
\label{matrixG}
\M{G}_{k}\,=\,\sum\limits_{p=0}^{k-1} \M{D}^{p}
\,=\,\left(\M{1}-\M{D}\right)^{-1}\left(\M{1}-\M{D}^{k}\right)
\quad\text{(using $\M{G}_{0}=\M{0}$).}
\end{equation}
(Obviously, the second equality from the left holds only as long as
$(\M{1}-\M{D})$ is invertible.)

$\M{S}^{N}$ has the same sparsity pattern, {\sl i.e.} the same
distribution of zeros, as $\M{S}$. In particular, its upper
left $2\times 2$ submatrix reads
$\left(\begin{smallmatrix} 1 & \tilde{a} \\ 0 & 1\end{smallmatrix}\right)$,
thus preserving the right eigenvector $\trans{(1,0,0,0)}$, standing
for the invariance to a global voltage shift, and the left eigenvector
$(0,1,0,0)$, responsible for current conservation, both with
eigenvalue $\mu=1$.

The remaining two eigenvalues are $\mu$ and $1/\mu$ (since
$\det\M{S}^{N}=\det\M{S}=1$), and have their origin in the lower right
submatrix, $\M{D}^{N}$. Using
$\M{D}^{2}=\M{D}\,\trace{\M{D}} - \M{1}\det{\M{D}}$
(valid for any $2\times 2$ matrix), together with $\det{\M{D}}=1$, we
have
\begin{subequations}
\begin{equation}
\label{matrixDNa}
\M{D}^{N}\;=\;\M{D}^{N-2}\M{D}^{2}\;=\;2\xi\,\M{D}^{N-1}-\M{D}^{N-2}
\;\text{,}
\end{equation}
with $\xi=\frac{1}{2}\trace{\M{D}}$. The recurrence relation
(\ref{matrixDNa}) may be thought of as a matrix version of the one
defining the Chebyshev polynomials, eq.~(\ref{chebyshev}). It may be
easily verified that
\begin{equation}
\label{matrixDNb}
\M{D}^{N}\;=\;U_{N-1}(\xi)\M{D}\,-\,U_{N-2}(\xi)\M{1}
\;\text{,}
\end{equation}
\end{subequations}
(with $U_{n}(\xi)$ the Chebyshev polynomials of the second kind,
eq.~(\ref{chebyU})) fulfils eq.~(\ref{matrixDNa}) and meets the initial
conditions $\M{D}^{0}=\M{1}$ and $\M{D}^{1}=\M{D}$. Quite similarly,
for the trace of $\M{D}^{N}$, (\ref{matrixDNa}) gives
\begin{subequations}
\begin{equation}
\label{traceDNa}
\trace{\M{D}^{N}}\;=\;2\xi\,\trace{\M{D}^{N-1}}-\trace{\M{D}^{N-2}}
\end{equation}
which, along with the initial conditions
$\trace{\M{D}^{0}}=\trace{\M{1}}=2$ and $\trace{\M{D}^{1}}=2\xi$,
yields
\begin{equation}
\label{traceDNb}
\trace{\M{D}^{N}}\;=\;2 T_{N}(\xi=\cos\theta)\
\;\equiv\;2\cos\left[N\theta\right]
\;\text{,}
\end{equation}
\end{subequations}
where $T_{N}$ are the Chebyshev polynomials of the first kind.

Since the trace of a matrix is invariant under basis transformation,
the resonance condition, $\mu=1$, may be reformulated as
$\trace{\M{D}^{N}}=\mu+1/\mu=2$, insertion of which in
eq.~(\ref{traceDNb}) requires $\theta_{m}=\frac{2\pi m}{N}$ with
$m=0\ldots N-1$, and thus finally:
\begin{equation}
\label{traceresonance}
\frac{1}{2}\trace{\M{D}}=\xi=\cos\theta_{m}=\cos\frac{2\pi m}{N}
\end{equation}
The resonance condition (\ref{traceresonance}) is the main result of
this section; it states that, in order to compute all resonances of a
chain of $N$ quadrupoles under periodic boundary conditions, one
simply has to calculate the transfer matrix $\M{S}$ of a single
quadrupole --- which, of course, depends on the components constituting
the quadrupole and their setup --- and then resolve the algebraic
equation (\ref{traceresonance}) for the lower right $2\times 2$
submatrix $\M{D}$.

\subsection{Applications}

To see this recipe at work, let us recalculate the resonances of an
array obtained by replication of the unit cell of
Fig.~\ref{fig:unitcells}(c). Using the capacitors' conductance as a
reference, $\sigma_{0}=\sigma_{C}=\ic\w C$, the quadrupole's transfer
matrix $\M{T}$ may be written (from right to left) as a product of the
transfer matrices of a vertical capacitor, $\M{T}_{\rm B}(2)$ (with
twice the capacity $C$, due to the periodic boundary conditions along
the electrodes), followed by an inductor and a capacitor in parallel,
$\M{T}_{\rm A}(\frac{\sigma_{L}}{\sigma_{C}},1)$, with
$\sigma_{L}/\sigma_{C}=-1/(LC\w^2)=-1/\wt^2$, another vertical
capacitor, and finally two capacitors in parallel, $\M{T}_{\rm
A}(1,1)$:
\begin{eqnarray}
\label{matrixSunitcellc}
\M{S} &=&
\M{P}\, 
\M{T}_{\rm A}(1,1)\,
\M{T}_{\rm B}(2)\,
\M{T}_{\rm A}(\sfrac{\sigma_{L}}{\sigma_{C}},1)\,
\M{T}_{\rm B}(2)\,
\M{P}
\nonumber\\
&=&
\begin{pmatrix}
1 \;&-\frac{3}{2}+\frac{1}{2}\wt^2 \;& -2+2\wt^2  \;& \frac{1}{2}+\frac{1}{2}\wt^2\\
0 \;& 1                            \;& 0          \;& 0                           \\
0 \;& \frac{5}{2}+\frac{5}{2}\wt^2 \;& 19-10\wt^2 \;&-\frac{7}{2}+\frac{5}{2}\wt^2\\
0 \;& -2-2\wt^2                    \;& -16+8\wt^2 \;& 3-2\wt^2
\end{pmatrix}
\end{eqnarray}
A straightforward calculation shows that, in this case, the resonance
condition (\ref{traceresonance}) reduces to
\begin{equation}
\label{dispersLC1period}
\wt_{m}=\pm \sqrt{\frac{11}{6}-\frac{1}{6}\cos\frac{2\pi m}{N}}
=\pm \sqrt{\frac{5}{3}+\frac{1}{3}\sin^{2}\frac{m\pi}{N}}
\;\text{,}
\end{equation}
with $m=0\ldots N-1$. The only difference between the eigenfrequencies
(\ref{dispersLC1period}), calculated for periodic boundary conditions
in both, $x$- and $y$-direction, and the former result,
eq.~(\ref{dispersLC1}), obtained for a closed chain with periodic
boundary conditions only in $y$-direction, resides in the argument of
the sine function whose symmetry about $\frac{\pi}{2}$ organises the
eigenfrequencies (\ref{dispersLC1period}) in degenerate pairs (except
for $\wt_{0}$ and, for even $N$, $\wt_{N/2}$).

The second example to be analysed with the present method is the array
of staircases shown in Fig.~\ref{fig:array}, and obtained from the
unit cell in Fig.~\ref{fig:unitcells}(d). For its transfer matrix,
\begin{equation}
\label{matrixSunitcelld}
\M{S} \,=\,
\M{P}\, 
\M{T}_{\rm A}(\sfrac{\sigma_{L}}{\sigma_{C}},1)\,
\M{T}_{\rm B}(1+\sfrac{\sigma_{L}}{\sigma_{C}})\,
\M{T}_{\rm A}(1,\sfrac{\sigma_{L}}{\sigma_{C}})\,
\M{T}_{\rm B}(1+\sfrac{\sigma_{L}}{\sigma_{C}})\,
\M{P}
\text{,}
\end{equation}
the resonance condition (\ref{traceresonance}) is a fourth order
equation in $\wt^2$,
\begin{eqnarray*}
\wt^8-8\wt^6+\left(18-4\beta_{m}^2\right)\wt^4-8\wt^2+1&=& 
\\
\left[\wt^4-\left(4+2\beta_{m}\right)\wt^2+1\right]
\,
\left[\wt^4-\left(4-2\beta_{m}\right)\wt^2+1\right]&=&0
\,\text{,}
\end{eqnarray*}
with $\beta_{m}=\pm\cos\frac{\pi m}{N}$; its solutions are
\begin{equation}
\label{dispersstairs}
\wt_{m}^{2}\,=\,\beta_{m}+2\,\pm\,\sqrt{\left(\beta_{m}+2\right)^2-1}
\;\text{,}
\end{equation}
(where, again, $m\in[0,N/2]$ suffices, since the rest of the
resonance frequencies is obtained by symmetry).

\begin{figure}[htb]
\begin{center}
\includegraphics[width=0.8\cwidth]{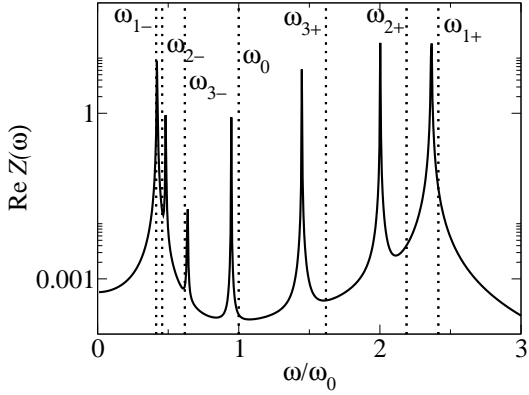}
\end{center}
\caption{\label{fig:renormstairs} 
Impedance between the electrodes for the setup shown in
Fig.~\ref{fig:array}, with an array generated from the unit cell in
Fig.~\ref{fig:unitcells}(d), replicated 3 times in $x$-direction and
twice in $y$-direction, embedded in a 9 by 4 capacitor lattice. The
real part of the ENR impedance, calculated for $L=C=1$ and $R_{\rm
coil}=10^{-4}$, is plotted with continuous lines. The dotted lines
indicate the resonance frequencies obtained analytically from
eq.~(\ref{dispersstairs}) for $N=3$ and 
$m=0$ ($\wt_{0}=1$ (doubly degenerate) 
and $\wt_{1\pm}=\sqrt{3\pm 2\sqrt{2}}$),
and $m=1$ ($\wt_{2\pm}=\sqrt{{5}/{2}\pm{\sqrt{21}}/{2}}$ and
$\wt_{3\pm}=\sqrt{{3}/{2}\pm{\sqrt{5}}/{2}}$ }
\end{figure}
Fig.~\ref{fig:renormstairs} displays the impedance of the staircase
array of Fig.~\ref{fig:array}, for which $N=3$. A comparison of the
resonance frequencies (\ref{dispersstairs}) (dotted lines) with an ENR
calculation for the same array (continuous lines) shows that
eq.~(\ref{dispersstairs}) not only predicts the right number of
resonances --- this result is non-trivial since, without degeneracy,
eq.~(\ref{dispersstairs}) would produce 8 resonances ( 4 for $m=0$ and
another 4 for $m=1$) instead of 7 --- but also matches the position of
most of the resonances. The deviations, observed in particular for
$\wt_{2+}$ and $\wt_{3+}$, are due to the assumption of periodic
boundary conditions along the chain, an approximation used in the
derivation of eq.~(\ref{traceresonance}).

\subsection{Link to the usual resonance condition}

Up to now, the resonance condition (\ref{traceresonance}) had to be
thought of as a necessary one, since $\mu=1$ for all four eigenvalues
of $\M{S}^{N}$ was merely required for the norm of the eigenvectors
$\V{J}$ to remain finite after a great number of transfers through the
same chain of $N$ quadrupoles. The aim of this paragraph is to show
the equivalence of the new resonance condition (\ref{traceresonance})
with the more familiar one, namely a vanishing total current through
any cross-section of the chain,
$I_{k}^{+}=\frac{1}{\sqrt{2}}(I_{k}+I_{k'})=0$, despite finite applied
voltage.

At resonance, $\M{D}^{N}$ may be evaluated explicitly: from
eq.~(\ref{traceresonance}), $\theta_{m}=\frac{2\pi m}{N}$, the
recurrence relation (\ref{matrixDNb}), and the Chebyshev polynomials
(\ref{chebyU}) evaluated at resonance,
\begin{align}
\label{chebyUres}
U_{N-1}(\cos\theta_{m})&
=\frac{\sin\left[N\theta_{m}\right]}{\sin\theta_{m}}=
\begin{cases}
N & \text{if $m=0$,} \\
0 & \text{else}
\end{cases}
\nonumber
\\
U_{N-2}(\cos\theta_{m})&
=\frac{\sin\left[(N-1)\theta_{m}\right]}{\sin\theta_{m}}=
\begin{cases}
N-1 & \text{if $m=0$,} \\
-1  & \text{else}
\end{cases}
\nonumber
\end{align}
one obtains
\begin{equation}
\label{matrixDNres}
\M{D}^{N}=
\begin{cases} 
N\M{D}\,-\,\left(N-1\right)\M{1} & \text{if $m=0$,}\\
\M{1} & \text{else ({\sl i.e.} $m=1\ldots N-1$).}
\end{cases}
\end{equation}

For the excited eigenmodes, $m \mod N\neq 0$, $\M{D}^{N}=\M{1}$
implies $\M{G}_{m}=\M{0}$ (eq.~(\ref{matrixG})), and hence, from
eq.~(\ref{matrixSN}),
\begin{equation}
\label{matrixSNexcres}
\M{S}^{N}=
\begin{pmatrix}
\left(\begin{smallmatrix} 1 & \tilde{a} \\ 0 & 1 \end{smallmatrix}\right)
 &\;& \M{0} \\
\M{0} &\;& \M{1}
\end{pmatrix}
\text{.}
\end{equation}
Generally $\tilde{a}$ is finite; it is thus obvious that {\em any}
vector $\V{J}$ is a right eigenvector of $\M{S}^{N}$ with eigenvalue
$\mu=1$, if (and only if) its second component, $I^{+}$, describing
the net current through the sample, vanishes.

For the groundstate, $m=0$, the line of reasoning is slightly more
subtle: the naive procedure of solving the matrix equation
(\ref{naivetransfer}) at resonance, $I_{1}+I_{1'}=0$, amounts in the
present language to solving
\begin{equation}
\M{S}^{N}
\left(\begin{smallmatrix} U \\0 \\ x \\y \end{smallmatrix}\right)
\,=\,
\left(\begin{smallmatrix} U \\0 \\ x \\y \end{smallmatrix}\right)
\;\text{.}
\end{equation}
The right eigenvector $\trans{(U,0,0,0)}$ may be subtracted from the
system of equations, and one is left with an eigenvalue problem for
the last two components of $\V{J}$,
$\M{D}^{N}\left(\begin{smallmatrix} x \\y \end{smallmatrix}\right)
=\left(\begin{smallmatrix} x \\y \end{smallmatrix}\right)$, which ---
using eq.~(\ref{matrixDNres}) --- reduces to finding the right
eigenvector of the single quadrupole's $\M{D}$, associated with
eigenvalue $\mu=1$:
\begin{equation}
\M{D}
\left(\begin{smallmatrix} x \\y \end{smallmatrix}\right)
\,=\,
\left(\begin{smallmatrix} x \\y \end{smallmatrix}\right)
\end{equation}
(Note that, for the generally asymmetric matrix $\M{D}$, the right
eigenspace contains only one eigenvector if --- as in the present case
of a resonance --- the eigenvalues are degenerate.)

In all cases --- for the groundstate and for the excited modes --- the
condition that $\M{S}^{N}$ shall only have eigenvalues $\mu=1$ is
equivalent to the usual definition of a divergent impedance between
the electrodes, implying $I_{k}+I_{k'}=0$ at finite applied voltage.

\section{\label{sec:inflattice}Clusters in an infinite lattice}

%
%
\begin{figure*}[htb]
\begin{center}
\includegraphics[height=0.45\cwidth]{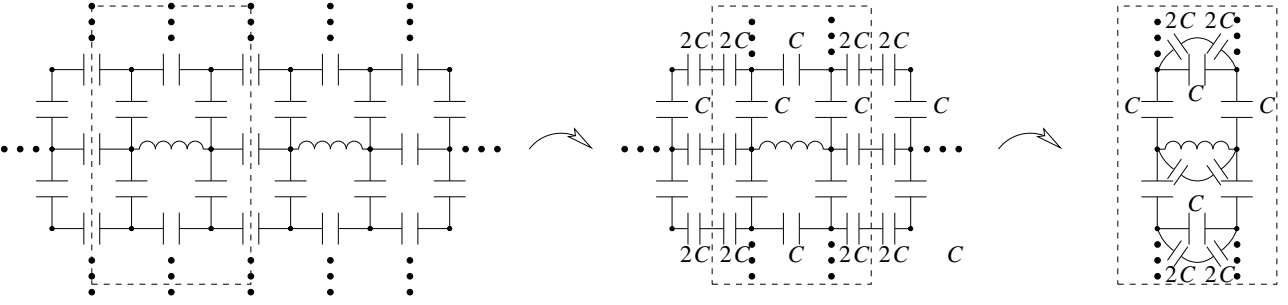}
\end{center}
\caption{\label{fig:LClineLF}Low-frequency renormalisation scheme for
the transformation of an infinite line of alternating coils and
capacitors in a 2D capacitor lattice with bonds of capacity $C$ (left
graph) to an infinite ladder of capacitors with only one horizontal
capacitor replaced by a coil (right graph). (In the second and third
graph, a bond with two capacitors in series {\em always} represents
two times $2C$ in series.)}
\end{figure*}

The systems studied in the preceding sections could be reduced to
effective 1D problems, because (i) the arrays covered the whole width
of the electrodes and (ii) periodic boundary conditions were assumed
in $y$-direction. In this section, by contrast, we will examine simple
regular 1D arrays in an infinite square lattice. The observed changes
turn out to be considerable, and sometimes not only shift the
frequencies, but qualitatively change the spectrum.

The first example we want to inspect consists in a single line of
alternating inductors and capacitors embedded in an otherwise pure
capacitor lattice, as shown in the leftmost graph of
Fig.~\ref{fig:LClineLF}. Alternatively, this line is generated by
replicating the unit cell of Fig.~\ref{fig:unitcells}(c) only in
$x$-direction.

\begin{figure}[htb]
\begin{center}
\includegraphics[width=0.95\cwidth]{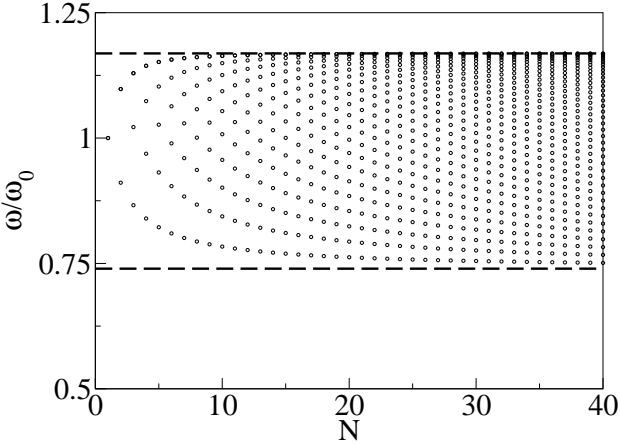}
\end{center}
\caption{\label{fig:LCworm}Eigenfrequencies $\w$ in units of $\wo$ for
a line of $N$ alternating coils and capacitors embedded in an infinite
capacitor lattice. The circles represent numerically calculated
eigenfrequencies, obtained with the spectral method. The dashed lines
are the analytical large $N$ asymptotes, $\wt_{\rm LF}\simeq
1/\sqrt{2\sqrt{2}-1}\simeq 0.7395$ and $\wt_{\rm HF}\simeq
1/\sqrt{\sqrt{3}-1}\simeq 1.1688$ discussed in the text.}
\end{figure}
Its eigenfrequencies are obtained very accurately with the so-called
spectral method, initially proposed by Straley~\cite{Straley},
Bergman~\cite{Bergman} and Milton~\cite{Milton}, and later adapted to
2D lattices by Clerc, Giraud, Luck and
co-workers~\cite{Clerc1996,Luck1998}, employing a Green's function
formalism developped by McCrea and Whipple~\cite{McCrea} in the
context of random walks; (see also Spitzer~\cite{Spitzer}).
This approach connects the impedance resonances of a (not necessarily
ordered) cluster to the eigenvalues of a non-symmetric square matrix,
which --- in general --- have to be evaluated numerically. The latter
version of the spectral method assumes the cluster to be located in
the middle of an infinite lattice --- just as in the examples we wish
to study. To obtain similar results from an ENR calculation one would
have to incorporate the cluster in a much larger capacitor network
(compared to the cluster's size) --- a situation which is
computationally very demanding.

The eigenfrequencies of the line of alternating coils and capacitors,
obtained via the spectral method, are plotted with dots in
Fig.~\ref{fig:LCworm} as a function of the number of coils, $N$. In
the limit of large $N$, the lowest eigenmode shows the full
translational invariance of the system, and thus has the same current
distribution in every vertical stripe. This implies that the left and
right boundary of the stripe can be thought of as attached
together. In order to perform this operation, illustrated in
Fig.~\ref{fig:LClineLF}, one has to (i) chose a stripe which is
symmetric about one of the coils, (ii) replace all horizontal
capacitors of capacity $C$ in all pure capacitor columns by two
capacitors in series, with capacity $2C$ each, (iii) cut between the
doubled capacitors, and (iv) tie the corresponding vacant ends of the
stripe together. One ends up with an infinite ladder with horizontal
steps of $2C$ (two times $2C$ in series, the whole parallel to a
capacitor $C$), except for one step which contains a coil instead of
the single capacitor; the vertical legs of the ladder consist of
capacitors $C$ ({\sl cf.} rightmost graph in Fig.~\ref{fig:LClineLF}).
The capacities may then be summed up in a procedure analogous to
eqs.~(\ref{LCinf}) and (\ref{Cinf}), yielding a total capacity of
$C(2\sqrt{2}-1)$, and thus the low-frequency asymptote $\wt_{\rm
LF}\simeq 1/\sqrt{2\sqrt{2}-1}\simeq 0.7395$ (lower dashed line in
Fig.~\ref{fig:LCworm}).

The position of the high-frequency asymptote $\wt_{\rm HF}\simeq
1/\sqrt{\sqrt{3}-1}\simeq 1.1688$ (upper dashed line in
Fig.~\ref{fig:LCworm}), on the opposite side of the spectrum, can be
calculated almost analogously: in the limit of an infinite $LC$ chain
in its highest eigenmode, the currents through two neighbouring coils
are at any moment antiparallel, but of the same magnitude. The
capacitors separating the coils horizontally are thus located at
current nodes and can be omitted. The corresponding ladder looks the
same as the one shown in the right of Fig.~\ref{fig:LClineLF}, except
for the fact that all capacitors of $2C$ (on the bent lines) have to
be removed.

Comparison to the case with periodic boundaries in $y$ direction,
studied in Sec.~\ref{sec:kirchhoff} (eq.~(\ref{dispersLC1}) in
particular), shows that the shift of a resonance depends on its
location in the spectrum: at the low-frequency edge, formerly located
at $\wt\simeq\sqrt{5/3}\simeq 1.291$, now at $\wt\simeq 0.7395$, the
renormalisation is stronger than at the high-frequency boundary
(formerly $\wt\simeq\sqrt{2}\simeq 1.414$, now $\wt\simeq
1.1688$). The reason for this behaviour lies in the different
renormalisation schemes which --- as pointed out above --- couple the
inductors to an equivalent capacity which is less enhanced for high
frequencies than for low ones.

The second cluster we are going to study in this section is a straight
line of $N$ coils in series embedded in a pure capacitor environment.
\begin{figure}[htb]
\begin{center}
\includegraphics[width=0.95\cwidth]{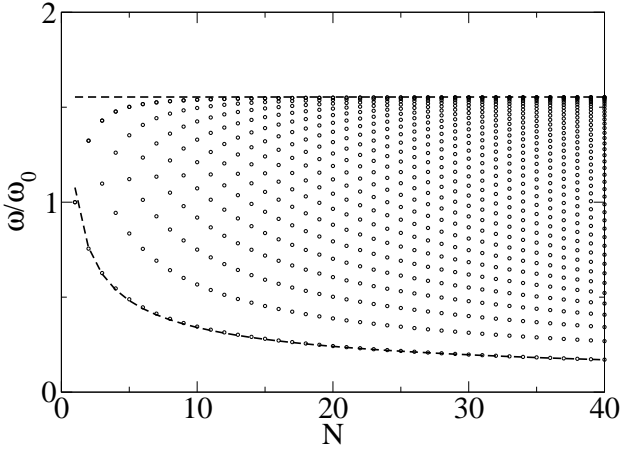}
\end{center}
\caption{\label{fig:Lworm}Eigenfrequencies $\w$ in units of $\wo$ for
a continuous line of $N$ coils embedded in an infinite capacitor
lattice. The circles represent numerically calculated
eigenfrequencies, obtained with the spectral method. The dashed lines
are the large $N$ asymptotes, $\wt_{\rm LF}\simeq\sqrt{1.16/N}$ and
$\wt_{\rm HF}\simeq\sqrt{1+\sqrt{2}}\simeq 1.5538$ (see text).}
\end{figure}
The resonance frequencies of this ``lattice worm'', obtained via the
spectral method, are plotted with dots in Fig.~\ref{fig:Lworm}. For
large $N$ and at high frequencies, the cluster can again be reduced to
almost independent stripes.
%
%
\begin{figure*}[htb]
\begin{center}
\includegraphics[height=0.45\cwidth]{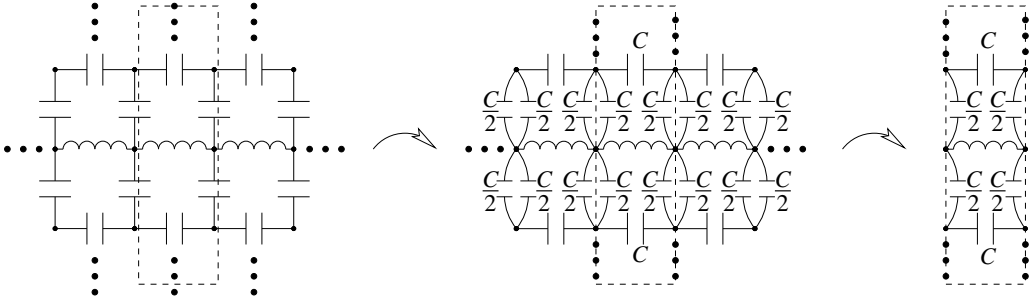}
\end{center}
\caption{\label{fig:LlineHF}High-frequency renormalisation scheme for
an infinite line of coils in a 2D capacitor lattice (left graph),
ending up in an infinite capacitor ladder with only one coil (right
graph).}
\end{figure*}
The corresponding renormalisation scheme is illustrated in
Fig.~\ref{fig:LlineHF}, and ends up in an infinite ladder with legs of
capacity $C/2$ and steps of capacity $C$ (except for the central step
which is a coil of inductance $L$). Summing up each semi-infinite
capacitor ladder yields
$C_{\rm inf}=\frac{1}{2}C(1+\sqrt{2})$
({\sl cf.} eqs~(\ref{LCinf}) and (\ref{Cinf})), and thus the
high-frequency asymptote 
$\wt_{\rm HF}\simeq \sqrt{1+\sqrt{2}}\simeq 1.5538$
(upper dashed line in Fig.~\ref{fig:Lworm}). Comparison to the
periodic case, eq.~(\ref{dispersL2}), where $\wt\to\sqrt{3}$ for $m=N$
and $N\to\infty$, shows that the renormalisation shifts are rather
modest at high frequencies.

At low frequencies, by contrast, the infinite environment has drastic
effects: as can be seen from Fig.~\ref{fig:Lworm}, the lowest
resonance tends to zero frequency as the length of the worm
increases. The corresponding density of states therefore shows no gap
at low frequencies --- as opposed to the density of states
(\ref{dosL2}) of the periodic system, illustrated in the last graph of
Fig.~\ref{fig:dos}.

The reason for this behaviour can be qualitatively understood in a
simple picture representing the whole worm of $N$ coils as a dipole of
total inductance $N L$. This dipole is thought to be coupled to a
single capacitor bearing the entire capacity of the lattice. In this
model, the lowest resonance would thus be found at
$\w\simeq\sqrt{1/(N L C_{\rm lat})}$.
As pointed out in Appendix~\ref{app:dipole}, the dipole model
evaluates the entire lattice capacity to roughly $C_{\rm lat}\simeq
C$, a value which allows to reproduce the exact resonance frequency
$\wt=1$ for a single coil ($N=1$) in a 2D capacitor lattice. For
increasing $N$, $C_{\rm lat}$ is found slightly reduced since more and
more capacitors are replaced by coils. In the limit of large $N$, the
numerically calculated groundstate fits very accurately to
$\wt\simeq\sqrt{1.16/N}$ (lower dashed line in Fig.~\ref{fig:Lworm}).

In the same manner, the dipole model predicts a resonance frequency of
$\wt_{m}\propto\sqrt{m/N}$ for the $m$.th mode, with a proportionality
constant of the order of unity, since the current nodes split the
``worm'' into $m$ almost independent segments of $N/m$ coils
each. Comparison to the numerically evaluated eigenfrequencies shows
that this scenario works well for $m\lesssim 5$, but turns out to be
too simplistic for the higher excited modes.

On more general grounds, and observing that second-order terms do not
improve the reproduction of the data, one may try an ansatz of the
form
\begin{equation}
\label{dispersLfit}
\wt_{m}\;\simeq\;\sqrt{\frac{\alpha_{m}}{N}}\;+\;\sqrt{\frac{\beta_{m}}{N}}^{3}
\;\text{.}
\end{equation}
A linear dependence of the parameters, $\alpha_{m}=-0.40+1.57 m$ and
$\beta_{m}=-0.56+0.98 m$, deduced by a fit for $N=200\ldots 1000$ and
$m=1\ldots 30$, reproduces the numerically calculated resonance
frequencies very accurately for a wide range of $N$ and $m$, as long
as $N/m$ is large ($\gtrsim 5$).

We also note that the fitted $\alpha$, with a slope of $1.57$,
concords with an analytical result obtained by Clerc {\sl et al.}
\cite{Clerc1996}, stating that $\wt_{m}\sim\sqrt{\frac{\pi}{2} m/N}$ for
the bulk states of a long linear cluster, {\sl i.e.} if both $m$ and
$N$ are large.

\section{\label{sec:conclusions}Conclusions and outlook}

In this article, we have studied the dielectric resonance spectra of
ordered passive arrays, typically --- although not necessarily ---
constituted of inductive and capacitive elements. Similar arrays,
based on split-ring resonators, have recently been used to assemble
metamaterials for the microwave regime, exhibiting negative refraction
and other exotic properties~\cite{Pendry2004,Seddon2003,Grbic2004}.

In the first part, we have calculated the resonance frequencies of
several arrays by solving a system of differential equations deduced
directly from Kirchhoff's rules --- a technique which allows for a
straightforward interpretation of the spectra and thus provides a good
handle for the influence of each of the array's parameters. On the
other hand, each new circuit requires a separate individual analysis,
relying on the solution of increasingly complicated recurrence
relations as the unit cell of the array gets richer in structure.

An alternative approach, presented in Sec.~\ref{sec:transfer}, deduces
the resonance frequencies from the array's transfer matrix, {\sl i.e.}
a matrix connecting the state vectors (a combination of currents and
voltages at each vertex) at both ends of the cluster to each
other. Within this formalism, the array is shown to be in resonance if
all eigenvalues of its transfer matrix are unity for a given frequency
--- a condition which is demonstrated to be equivalent to the more
familiar definition of a divergent impedance for a cluster between two
electrodes, namely vanishing net current through the sample at finite
applied voltage. Even large arrays, with complex unit cells, can be
easily analysed with this algorithm since, in a handier reformulation,
the new resonance condition, eq.~(\ref{traceresonance}), does not
require the computation of the transfer matrix of the whole array, but
only of a single unit cell. The latter is most conveniently evaluated
by multiplication of the transfer matrices of a few standard
situations --- two in our case, {\sl cf.}  eq.~(\ref{transferABgraph}),
where the 2D arrays reduce to two-legged ladders due to the assumption
of periodic boundary conditions along the electrodes --- a task which
can be performed using symbolic computation software. In its final
form, the resonance frequencies are typically given as the roots of a
polynomial whose degree depends on the complexity of the unit cell.

If, however, the cluster happens to be embedded in a much larger
lattice, periodic boundary conditions are not a pertinent
approximation, and the resonance frequencies are generically not
available in closed form anymore. In this case, major changes in the
spectrum have to be expected, of which some can be understood in terms
of renormalisation, while others cannot. Among the latter, one may
recall the example of a longer growing linear array: embedded in a
periodic medium, the spectrum remains gapped for any length of the
cluster; if, by contrast, the same cluster is isolated in a
homogeneous network, the gap closes with increasing length and the
resonance spectrum carries the signature of one or more rather
independent two-dimensional dipoles.

The tools discussed in this article may be useful for the design of
circuits with a custom electric response. Such devices could be used
for labelling --- just as bar codes --- which could be read by an
automated system.

For the future it would be undoubtedly desirable, especially in the
perspective of the already mentioned development of metamaterials with
negative refraction, to take into account the magnetic part of the
optical response, with the aim to create devices having both, tunable
dielectric {\em and} magnetic resonances.

\begin{acknowledgement}
We thank J.-M. Luck for very interesting discussions and remarks.
\end{acknowledgement}

\begin{appendix}
\section{\label{app:dipole} Dipole approximation for linear clusters}

\begin{figure}[htb]
\begin{center}
\includegraphics[width=0.95\cwidth]{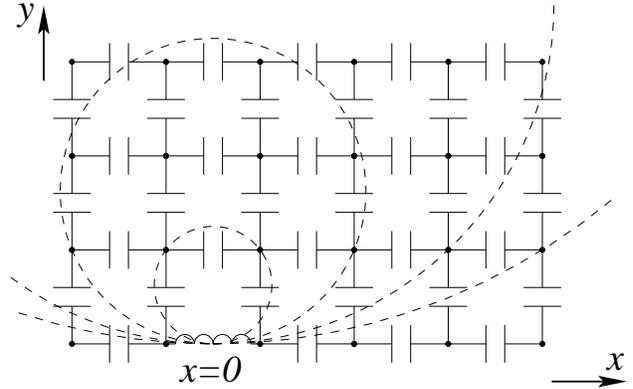}
\end{center}
\caption{\label{fig:dipolefield} Current distribution (dashed lines)
of a linear cluster of length $N=1$ in the dipole approximation. (The
lower half-plane has been omitted for clarity.)}
\end{figure}

The scenario presented in the following is based on the idea that the
current flowing through a linear cluster isolated in a capacitor
lattice flows back through the lattice with a current distribution
which, in the lowest eigenmode, resembles the field lines of a dipole
in 2D electrostatics (see Fig.~\ref{fig:dipolefield}).

This idea is supported by the spectral method, in the framework of
which the electric potential $V_{\V{r}}$ induced by a single
horizontal coil, located at the origin, is known to be given by
\cite{Clerc1996}
\begin{equation}
\label{Clerc2.12}
\lambda V_{\V{r}}\;=\; 
-W\,
\left[ 
G_{\V{r}+\frac{1}{2}\V{\hat{x}}}- G_{\V{r}-\frac{1}{2}\V{\hat{x}}} 
\right]
\;\text{,}
\end{equation}
where $\lambda$ is a frequency-dependent proportionality constant, $W$
the voltage measured between the ends of the coil, and $G_{\V{r}}$ the
2D lattice Green's function. Far from the coil, for $\abs{\V{r}}\gg
1$, the lattice Green's function may be replaced by its continuous
counterpart, $G_{\V{r}}\simeq -\frac{1}{2\pi}\ln\abs{\V{r}}$, and
eq.~(\ref{Clerc2.12}) reduces to a 2D dipole potential
\begin{equation}
\label{dipolepotential}
\lambda V_{\V{r}}
\;\simeq\;-W \frac{\partial}{\partial x} G_{\V{r}}
\;\simeq\;\frac{W}{2\pi}\;\frac{x}{x^{2}+y^{2}}
\;\equiv\;\frac{W}{2\pi}\;\frac{\cos\varphi}{\abs{\V{r}}}
\;\text{.}
\end{equation}

In this approximation, the current distribution follows the field
lines
\begin{equation}
\label{dipolefield}
r(\varphi)\;=\;\frac{l}{\pi}{\sin{\varphi}}
\end{equation}
where $l$ is the length of the field line (in units of the lattice
spacing). In order to calculate the resonance frequency of a cluster
embedded in an infinite capacitor lattice, one has to assign an
equivalent capacity, $C_{\rm lat}$, to the entire lattice. Of course,
$C_{\rm lat}$ depends on the current distribution: in general,
capacitors almost perpendicular to the current lines contribute only
little to $C_{\rm lat}$, while capacitors following the current lines
contribute as if they were in series. At great distance from the
cluster, the lattice behaves as if it were assembled by independent
lines of capacitors, hence all in parallel, contributing each as the
inverse of its length:
\begin{equation}
\label{Clat}
C_{\rm lat}\;\simeq\;\sum\limits_{k}\frac{C}{l_{k}}
\end{equation}

The sum in eq.~(\ref{Clat}) runs over all possible current paths, and
--- in order to avoid divergences --- the counting has to be done very
carefully. In our example of a single coil in an infinite lattice,
this may be achieved by supposing that the current can only leave the
$x$-axis at spots where a vertical capacitor is present. Following the
illustration of Fig.~\ref{fig:dipolefield}, this amounts to retaining
only current lines passing through $(x=k+\frac{1}{2},y=1)$ with
$k=0,1,2,\ldots$. In coordinates of the field lines
(\ref{dipolefield}), these points are described by
\begin{subequations}
\label{pinning}
\begin{eqnarray}
\label{pinningl}
l_{k}&=&\pi\left[1+\left(k+\sfrac{1}{2}\right)^2\right] \\
\label{pinningphi}
\sin\varphi_{k}&=&\frac{1}{\sqrt{1+\left(k+\frac{1}{2}\right)^2}}
\;\text{.}
\end{eqnarray}
\end{subequations}
The first equation can be used to select only the desired current
lines, as shown in Fig.~\ref{fig:dipolefield}. Pinning the field lines
in this way has a twofold virtue, namely on the short side to set a
lower boundary for the path length $l_{k}$ in (\ref{Clat}), {\em and}
to avoid the logarithmic divergence on the long side,
$l_{k}\to\infty$. After substitution in (\ref{Clat}), and taking into
account contributions from the lower half-plane, we obtain
\begin{eqnarray}
\label{Clatdip}
C_{\rm lat}
&\simeq& 
\frac{2 C}{\pi} \sum\limits_{k=0}^{\infty} \frac{1}{1+(k+\frac{1}{2})^{2}}
\nonumber \\
&\simeq& 
\frac{2 C}{\pi} \int\limits_{0}^{\infty} \frac{{\rm d}k}{1+k^{2}}
\;=\; C
\;\text{.}
\end{eqnarray}
(The sum in the first line may be thought of as a midpoint trapeze
approximation for the integral in the second line).

According to the dipole approximation, an isolated coil in a 2D
capacitor network thus resonates at
$\w=1/\sqrt{L C_{\rm lat}}\simeq 1/\sqrt{L C}\equiv\wo$ ---
which is the exact result known from the spectral method.

\end{appendix}


\end{document}